\documentclass[aps,prl,twocolumn,superscriptaddress,floatfix]{revtex4-1}

\usepackage{graphicx}
\usepackage{float}
 \usepackage{times}
\begin{document}

\title{Observation of Efimov molecules created from a resonantly interacting Bose gas}

\author{Catherine E. Klauss}
\affiliation{JILA, National Institute of Standards and Technology and University of Colorado, and Department of Physics, Boulder, CO 80309-0440, USA}
\author{ Xin Xie}
\affiliation{JILA, National Institute of Standards and Technology and University of Colorado, and Department of Physics, Boulder, CO 80309-0440, USA}
\author{Carlos Lopez-Abadia}
\affiliation{JILA, National Institute of Standards and Technology and University of Colorado, and Department of Physics, Boulder, CO 80309-0440, USA}
\author{Jos\'{e} P. D'Incao}
\affiliation{JILA, National Institute of Standards and Technology and University of Colorado, and Department of Physics, Boulder, CO 80309-0440, USA}
\author{Zoran Hadzibabic}
\affiliation{Cavendish Laboratory, University of Cambridge, J. J. Thomson Avenue, Cambridge CB3 0HE, UK.}
\author{Deborah S. Jin}
\affiliation{JILA, National Institute of Standards and Technology and University of Colorado, and Department of Physics, Boulder, CO 80309-0440, USA}
\author{Eric A. Cornell}
\affiliation{JILA, National Institute of Standards and Technology and University of Colorado, and Department of Physics, Boulder, CO 80309-0440, USA}

\date{\today}
\begin{abstract}
We convert a strongly interacting ultracold Bose gas into a mixture of atoms and molecules by sweeping the interactions from resonant to weak. By analyzing the decay dynamics of the molecular gas, we show that in addition to Feshbach dimers it contains Efimov trimers. Typically around 8\% of the total atomic population is bound into trimers, identified by their density-independent lifetime of about 100~$\mu$s. The lifetime of the Feshbach dimers shows a density dependence due to inelastic atom-dimer collisions, in agreement with theoretical calculations. We also vary the density of the gas across a factor of 250 and investigate the corresponding atom loss rate at the interaction resonance.
\end{abstract}

\maketitle

Experiments with ultracold atomic gases provide access to a vast array of intriguing phenomena, in part because of magnetically tunable Feshbach resonances.
In particular, recent experimental \cite{rem2013lifetime, fletcher2013stability,  makotyn2014universal, eismann2016universal, fletcher2016two} and theoretical \cite{diederix2011ground, yin2013quench, sykes2014quenching, rancon2014equilibrating, laurent2014momentum, rossi2014monte, piatecki2014efimov, smith2014two, kira2015coherent, jiang2016long, yin2016quench} advances have made resonantly interacting Bose gases an exciting new research topic \cite{chevy2016strongly}.
Unlike their fermionic counterparts, strongly interacting Bose systems are profoundly influenced by three-body phenomena, and help us understand the progression from two- through few- to many-body physics.

At the Feshbach resonance the {\it s}-wave scattering length $a$ diverges, and in the case of zero density the Feshbach molecule state, also of size $a$, merges with the free-atom state.
This diatomic resonant scenario is the prelude for a set of exotic few-body phenomena, namely the Efimov effect.
Although the Feshbach molecular state is unbound at the resonance, there exists an infinite log-periodic series of Efimov three-body bound states \cite{efimov1971yad,efimov1973energy}.
At $1/a \rightarrow 0$ the size of the $p^{{\rm th}}$ Efimov state ($p = 0, 1, 2 ...$) is larger than the previous by a factor by 22.7, and its binding energy $E_{{\rm T}}^{(p)}$ smaller by a factor of $22.7^2$ \cite{braaten2006universality, wang2013ultracold}.

At finite density $n$ many-body effects complicate the physics.
The system has an additional length scale, the interparticle spacing $n^{-1/3}$, that may determine how few- and many-body interactions scale.
Many questions arise, such as: what are the structure, strength, length scale and dynamics of the two-, few- and many-body correlations?
What does it mean to have a two- or three-atom molecule when it is embedded in a gas with interparticle spacing comparable to the molecular size?

Both the ambiguous constitution of two- and three-body states in a many-body environment and the short-lived quasi-equilibrium of a resonantly interacting Bose gas \cite{makotyn2014universal} complicate experiments.
For these reasons, many experiments simplify matters by reducing interactions to a well-understood regime before imaging \cite{makotyn2014universal, rem2013lifetime, eismann2016universal, fletcher2013stability}.
This interaction sweep can preserve resonance fossils in the form of perceived loss \cite{rem2013lifetime, fletcher2013stability, eismann2016universal}, momentum generation \cite{makotyn2014universal}, and molecule formation.

In this letter, we create a mixture of $^{85}$Rb (free atoms), $^{85}$Rb$^{*}_2$ (Feshbach dimers), and $^{85}$Rb$^{*}_3$ (Efimov trimers) by sweeping a resonantly interacting degenerate Bose gas onto the molecular states in the weakly-interacting regime ($na^3 \ll 1$).
Our theory for the relevant Efimov trimer predicts binding energy values spectroscopically indistinguishable (within our technical limitations) from the Feshbach dimer binding energy, $E_{\rm b}$.
However, the calculated Efimov molecule lifetime is an order of magnitude shorter than the Feshbach dimer's, making loss rates an invaluable tool for distinguishing the population of each component.
In addition to detecting a population of Efimov trimers, we measure the density-dependent atom-dimer collision rates at finite interactions.
We also present measurements of the apparent atom loss over two orders of magnitude in density, covering a significant fraction of a full Efimov period, and discuss the potential of these measurements for probing the few- and many-body effects in the resonantly interacting Bose gas.


\begin{figure}[h]
\includegraphics[width=\linewidth]{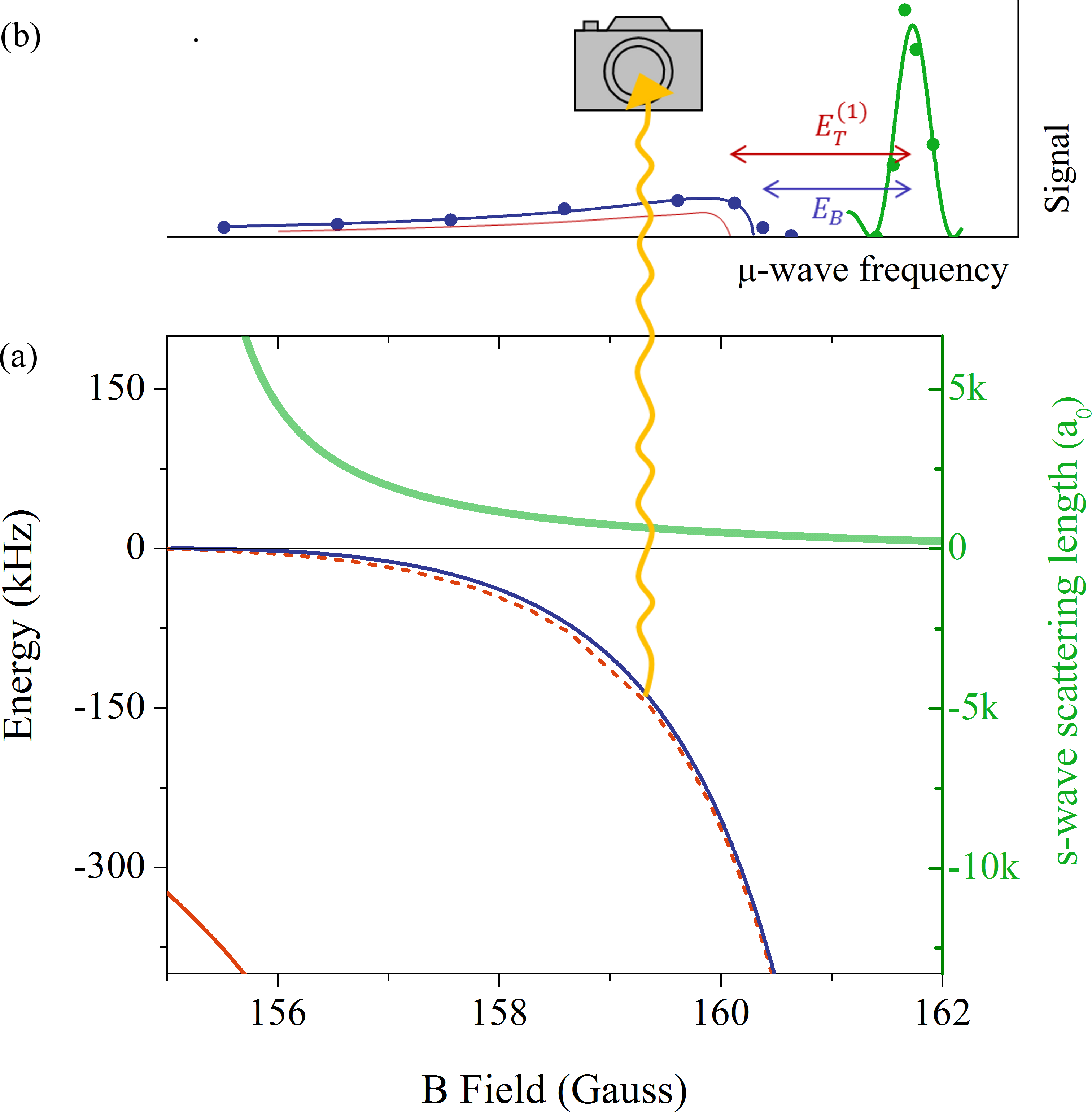}
\caption{(a) The solid green line shows the two-body scattering length near the Feshbach resonance centered at $155.04$ G.
The solid red line is the calculated energy of the ground state Efimov trimer, $E_{\rm T}^{(0)}$ \cite{mestrom2016efimov}.
The solid blue and dashed red lines show the calculated molecular energies of the Feshbach dimer, $E_{\rm b}$, and first excited Efimov trimer, $E_{{\rm T}}^{(1)}$, respectively.
We populate molecular states by sweeping our gas from 155.04 G to $B > 158.6$ G.
With a microwave pulse, we then transfer either atomic or molecular population to the $|3,-3\rangle$ state for absorption imaging; the experimental points in (b) are population transfer.
The green line is a delta function free-atom transition broadened by experimental resolution.
The blue line is a calculated Frank-Condon factor for molecular dissociation \cite{chin2005radio}.
The red line shows the expected shape of the trimer dissociation yield, but the difference between $E_{\rm T}^{(1)}$ and $E_{\rm b}$ is exaggerated for clarity.}
\label{fig:diagram}
\end{figure}

Our experiment begins with a $^{85}$Rb Bose-Einstein condensate (BEC) of $(6-8) \times 10^{4}$ atoms confined in a 10 Hz spherical magnetic trap in the  $|F, m_F \rangle = |2,-2 \rangle$ state \cite{pino2011photon}.
We use the Feshbach resonance centered at $B_0 = 155.041(18)$ G, with a width of $\Delta = 10.71(2)$ G, to control the interactions \cite{claussen2003very}.
We prepare our sample at $B \approx B_0 + 8$ G, at which $a \approx 150 \; a_0$, well within the extremely dilute limit with $ \langle n \rangle a^3 < 10^{-5} $.

At this point our condensate has a Thomas-Fermi density distribution with an average density of $ \langle n \rangle = 5.8(6) \times 10^{12}$ cm$^{-3}$.
We take advantage of tunable interaction strengths and our spherical trap to change to a mean density, $n_0$, before the main experimental sequence \cite{perez1997dynamics}.
$n_0$ defines our density-dependent scales for energy and time:
$E_n \equiv \hbar^2 (6\pi^2 n_{0})^{2/3} / 2m$ and $t_n \equiv \hbar/E_n$.

We rapidly lower the magnetic field from $B$ to $B_0$ in 5 $\mu$s following the procedure outlined in \cite{makotyn2014universal}.
We estimate that by the end of our ramp, the field is within 10 mG of the Feshbach resonance, which corresponds to $n_0 |a|^3> 10^{4}$, even for our lowest density.
After allowing the gas to evolve on resonance for $\tau_{\rm{evolve}} = 1.5 \; t_n$ we sweep to weak interactions ($n_0 a^3 < 0.001$) at a typical rate of 12.5 $\mu$s/G.
This sweeps the many-body resonant wave function onto the weakly-interacting free atomic state and the shallow molecular states, both diatomic and triatomic.
We measure the number of atoms swept into the molecular states by flipping one atom from each molecule from  $|2,-2\rangle$ to the $|3,-3\rangle $ imaging state, using microwave dissociation (see Fig. 1) \cite{chin2005radio}.
The 50 $\mu$s microwave pulse is detuned $>1.5 \times E_{\rm b}$ from the atomic resonance, and is long enough that all molecules are dissociated.

\begin{figure}
\includegraphics[width=\linewidth]{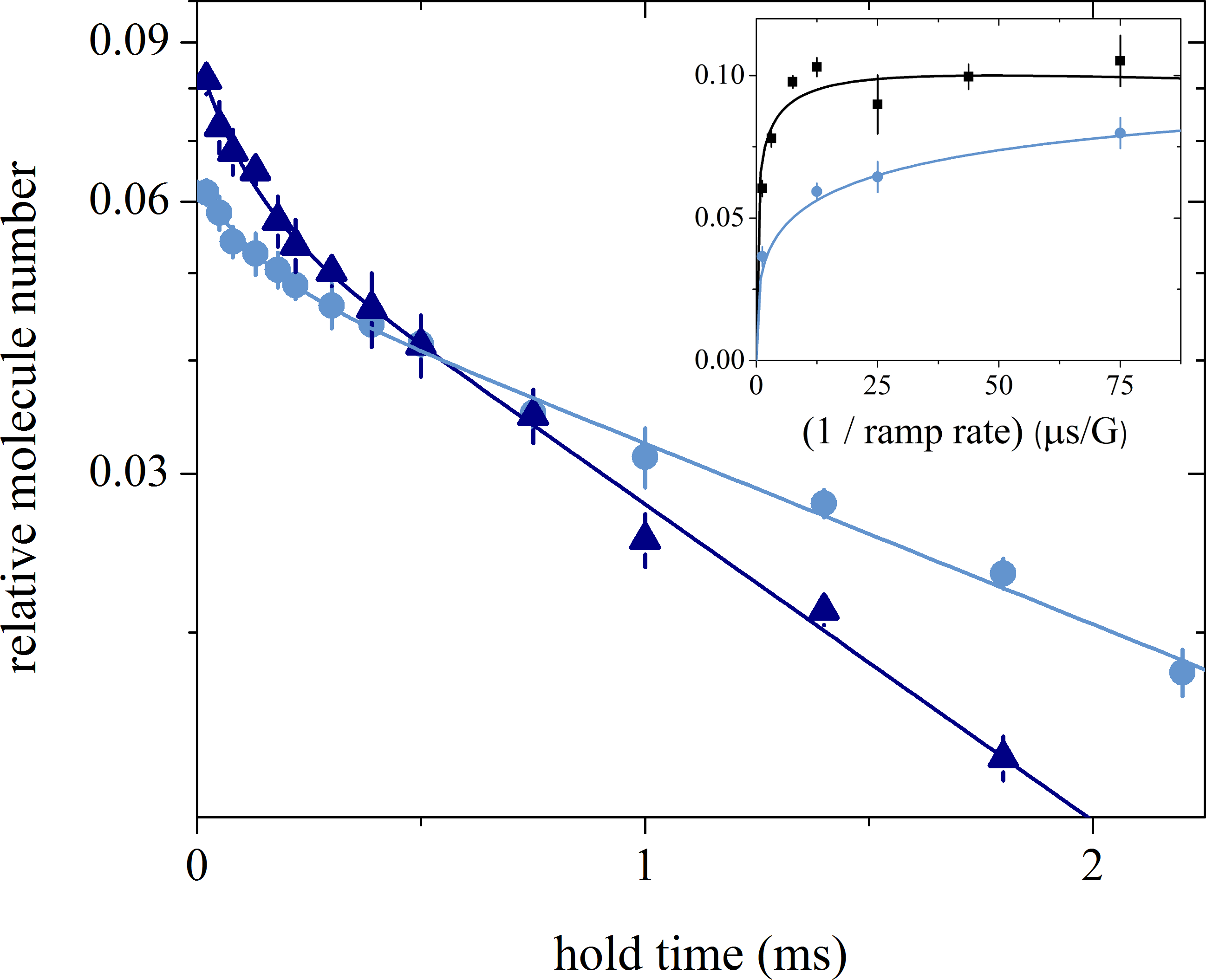}
\caption{
The number of molecules, normalized by the initial atom number, as a function of hold time at $a = 700(7)\,a_0$ for initial BEC densities $n_0 = 1.3(1) \times 10^{12}$ cm$^{-3}$ (dark triangles) and $n_0 = 0.20(1) \times 10^{12}$ cm$^{-3}$ (light circles).
The solid lines show a two component exponential decay.
At high and low densities the time for the initial fast loss  ($t_1$) does not significantly change, being 97(23) and 125(36) $\mu$s, respectively;
we identify this loss as decay of the trimers.
The time scales for the slower loss ($t_2$) are 1240(80) and 2200(100) $\mu$s;
as discussed in the text, this is consistent with a combination of dipole relaxation and collisional loss in the population of dimers, (see Eq. (3)).
For higher density data, the number of trimers is $A_1$ = 2000(200) and dimers $A_2$ = 4500(200).
This corresponds to 6000 atoms that were swept into the trimer state, roughly 8\% of the initial sample.
The inset shows the normalized number of molecules as a function of inverse ramp rate for $n_0 = 5.5(2) \times 10^{12}$ cm$^{-3}$ (dark squares) and $n_0 = 0.18 \times 10^{12}$ cm$^{-3}$ (light circles); the solid lines are guides for the eye.}
\label{fig:number}
\end{figure}

To investigate the composition of the molecular gas at weak interactions, we study its decay dynamics.
A typical measurement is shown in Fig. 2.
We find the data is described well by a sum of two exponentials: $A_1 e^{-t/t_1} + A_2 e^{-t/t_2}$, an indication of a two component mixture with distinct decay rates and populations of $A_1$ and $A_2$.

\begin{figure}
\includegraphics[width=\linewidth]{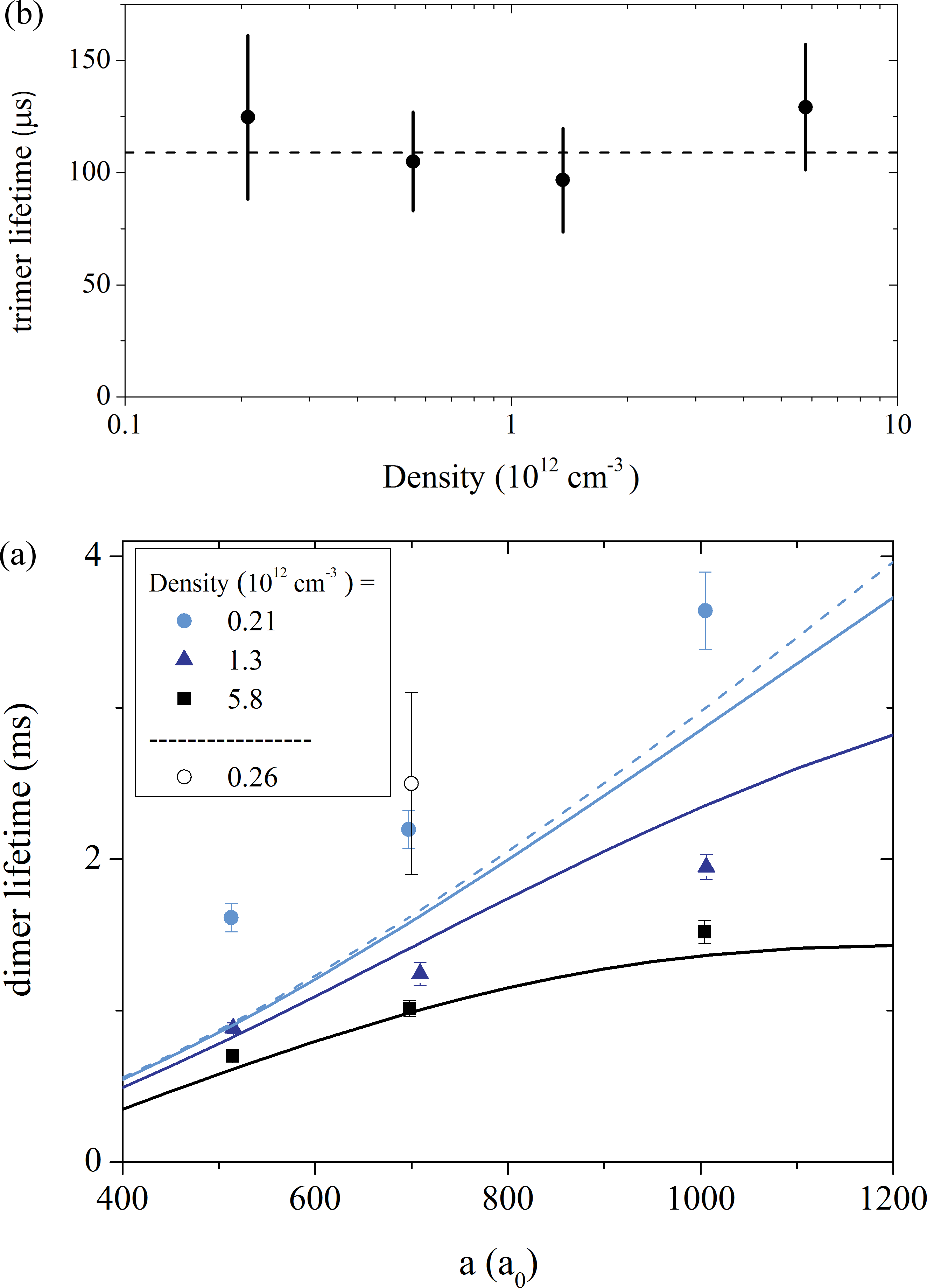}
\caption{(a) The lifetime of the Feshbach dimers for various densities as a function of the scattering length, $a$.
The solid lines represent the predicted dimer lifetime for the respective densities.
The dashed line represents the predicted dimer lifetime considering only spin relaxation.
While our lowest density molecules live somewhat longer than the inelastic spin relaxation model predicts, we see good agreement with previous measurements of dimer lifetimes from \cite{thompson2005spontaneous} for $n_0 = 0.26 \times 10^{12}$ cm$^{-3}$ (open circle).
(b) The lifetime of the Efimov trimers at $a = 700(7 )\; a_0$ average 114(16) $\mu$s (dashed line) over a factor of 30 in density.}
\label{fig:lifetimes}
\end{figure}

Figure 3(a) shows the fitted values of the longer lifetime, $t_2$, for various densities, and as a function of magnetic field.
For the lowest density ($n_0 = 0.21 \times 10^{12}$ cm$^{-3}$) our measurements are in qualitative agreement with theoretical predictions for inelastic spin relaxation in Feshbach dimers,
\begin{equation}
\tau_{D} = \tau_{\rm{res}} \frac{m a_{\rm bg}^2 \mu_{\rm{res}} \Delta}{2 \hbar^2} \left( \frac{B-B_0}{\Delta} \right) ^2 \left( \frac{\Delta}{B-B_0} -1 \right)^3,
\end{equation}
where $a_{\rm bg} = -443(3)\, a_0$ \cite{claussen2003very}, $\mu_{\rm{res}}/h = 34.66$ MHz/mT and $\tau_{\rm{res}} = 32\, \mu $s \cite{koehler2005spontaneous, comment1}.

Similar measurements at $\sim6$ and  $\sim28$ times larger densities yield shorter dimer lifetimes [Fig. 3(a)].
The density dependent loss rates are indicative of inelastic atom-dimer scattering: the collision of an atom with a shallow dimer produces a deeply bound molecule and a recoiling atom \cite{braaten2004enhanced, mestrom2016efimov}.
The dimer number is expected to decay roughly exponentially as this decay does not significantly deplete the much larger population of atoms.
The atom-dimer collisional rate coefficient $\beta$, defined by $\frac{d n_{\rm D}}{dt} = \frac{d n_{\rm A}}{dt} = - \beta n_{\rm A} n_{\rm D} $, is calculated in \cite{braaten2004enhanced} to be
\begin{equation}
\beta = \frac{20.3 \sinh(2\eta_{*})}{ \sin ^2 \left[ s_0 \ln(a/a_{*}) \right] + \sinh ^2 \left(\eta_{*}\right)} \frac{\hbar a}{m},
\end{equation}
where $\eta_{*}$ is the inelasticity parameter characterizing the width of the Efimov resonance at $a = a_{*}$, where the Efimov state intersects the atom-dimer threshold \cite{braaten2007resonant}.

The solid lines in Fig. 3(a) are theoretical predictions for the dimer lifetime due to inelastic spin relaxation and atom-dimer collisions:
\begin{equation}
\frac{1}{\tau_{\rm{Tot}}(B, n_0)} = \frac{1}{\tau_{\rm D}(B)} + \frac{1}{\tau_{{\rm AD}}(B, n')},
\end{equation}
where $\tau_{{\rm AD}} = \frac{1}{n' \, \beta}$ and we account for overall loss and modest expansion during $\tau_{\rm{evolve}}$ by setting $n' = 0.7 n_0$.
We use $\eta_{*} = 0.057$ \cite{wild2012measurements} and $a_{*} = 275\, a_0$ \cite{mestrom2016efimov}, and see qualitative agreement between our data and this model.
We therefore conclude the slower loss follows expected dimer loss rate patterns.

Returning now to explain the shorter lifetimes ($t_1$) seen in Fig. 2, we now discuss the other molecular states in this system: the infinite series of Efimov trimers that exist at $1/a \rightarrow 0$.
The Efimov ground state has an energy on the order of hundreds of kHz, [see Fig. 1(a)].
Of interest to our experiment is the first-excited Efimov state, whose energy $E_{\rm{T}}^{(1)}$ is on the order of hundreds of Hz \cite{wang2012origin, sykes2014quenching}, thus comparable to  $E_{n}$ in our experimental range of densities.
It is reasonable to expect that the mechanism that sweeps atoms into Feshbach dimers would also create Efimov trimers.
Although universal four-body states also exist \cite{von2009signatures, d2009universal}, for our experiment the first excited Efimov state is the only relevant weakly bound molecular state besides the Feshbach molecule \cite{comment2}.

We calculate ${E_{\rm T}}^{(1)}$ by solving the three-body problem in the adiabatic hyperspherical representation [see Fig. 1(a)] \cite{wang2012origin, sykes2014quenching}.
Our model, which includes a loss term tuned to match the Efimov decay seen in \cite{wild2012measurements}, shows that ${E_{\rm T}}^{(1)}$ is at most 7 kHz deeper than the Feshbach state and this difference varies only by 2 kHz between $500 \, a_0 $ and $1000 \, a_0$.
We calculate a lifetime of $109 \, \mu$s at $700 \, a_0$, determined by one-body decay to a deeply bound dimer state; this lifetime is an order magnitude shorter than the dimer lifetime, making the molecules distinguishable through lifetime alone.

Figure 3(b) shows the fitted values of $t_1$ at $a = 700(7) \, a_0$ for various initial densities.
The timescale of the fast decay does not appreciably vary over a range of 30 in density, and averages a value of 114(16) $\mu$s, in agreement with our prediction \cite{comment3}.
We therefore take this fast decay as evidence that the excited Efimov state has been populated.

Many experiments are sensitive to the existence of Efimov states, through both observation of inelastic collision rates in atomic samples \cite{kraemer2006evidence, zaccanti2009observation, wild2012measurements, berninger2011universality, gross2010nuclear, huckans2009three, barontini2009observation} and atom-dimer resonances \cite{ knoop2009observation, nakajima2010nonuniversal, zenesini2014resonant, bloom2013tests, lompe2010atom}, and observation of atomic loss after RF association into Efimov states \cite{lompe2010radio, machtey2012association, nakajima2011measurement}, and other means \cite{fletcher2016two}. This work, along with that in Ref. \cite{kunitski2015observation}, differs in that it is an observation of a {\it populated} Efimov state.

\begin{figure}
\includegraphics[width=\linewidth]{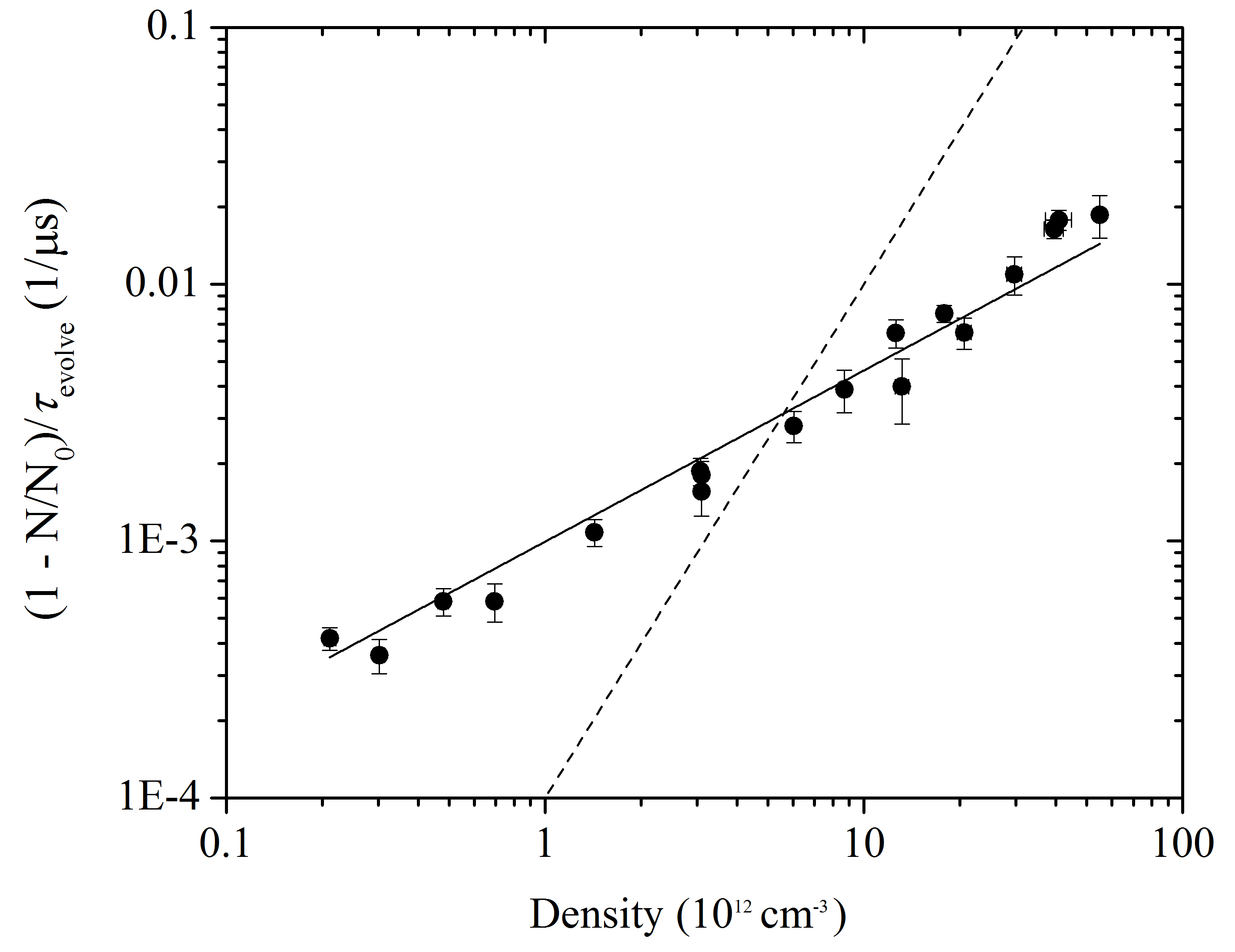}
\caption{$\left( 1 - N/N_0 \right)/\tau_{\rm{evolve}} $ is the perceived atom loss rate.
We measure the number of atoms that return to weak interactions after $\tau_{\rm{evolve}} \approx t_n$.
The solid line goes as $n^{2/3}$.
The dashed line goes as $n^2$, the expected three-body loss rate in the mean-field regime.}
\label{fig:atomloss}
\end{figure}

We find that the population of Efimov and Feshbach states depends on both the sweep rate (see inset of Fig. 2) and the evolution time on resonance, $\tau_{\rm evolve}$;
for a fast sweep and $\tau_{\rm evolve} \ll t_n$  we see almost the entire population return as free atoms.
Whether the atoms collisionally decay into deeply bound molecules, or are swept into weakly bound ones, their disappearance is a signature of the correlations that develop as the sample evolves at resonance.

What is the relevant physics that sets the rate at which these correlations appear?
The physical range of the two-atom interaction potential, $r_{\rm{vdw}}$, is likely too small to be relevant, whereas the scattering length $a$ and the initial thermal wavelength are likely too large.
What is left is only the mean interparticle spacing $n^{-1/3}$ and the sizes of the Efimov molecule, $R_{\rm{T}}^{(p)}$.
On resonance $R_{\rm{T}}^{(p)}$ is not experimentally adjustable, but by studying the rate of atom loss over a range in densities covering a factor of 250, we can change the ratio $R_{\rm{T}}^{(p)} / n^{-1/3}$ by a factor of 6.5, which is more than half of a complete log-periodic Efimov cycle.
To cover such a wide range of density, we perform a simplified protocol, in which we fix $\tau_{\rm{evolve}} \approx t_n$, sweep to weak interactions in 5 $\mu$s, and simply look at the total loss of the atomic population.
In this protocol, both recombination to deep states and conversion to shallow dimers and trimers look like atom loss.
In Fig. 4 we plot the atom-loss rate as a function of the initial density, and find that it agrees quite well with an $n^{2/3}$ power law.

To the extent that the atom loss is due to three-body recombination, the $n^{2/3}$ scaling makes sense: in the mean-field zero-temperature limit, the per atom three-body recombination rate goes as $n^2 a^4$ \cite{weber2003bose, nielsen1999low, esry1999recombination, bedaque2000three}; as $a$ formally diverges, a plausible physical limit is $a \sim n^{-1/3}$, yielding a loss rate scaling as $n^{2/3}$ \cite{d2004limits, chevy2016strongly}.
While it is known that at finite $a$ the presence of Efimov states modulates the three-body inelastic collision rates by a dimensionless log-periodic function of $a$ \cite{braaten2006universality, kraemer2006evidence, nielsen1999low, esry1999recombination, bedaque2000three, zaccanti2009observation, pollack2009universality, gross2009observation, ferlaino2011efimov, chin2010feshbach}, it is not known how this Efimovian physics modulates density dependence when both $T$ and $1/a \rightarrow 0$.

As for the other contributing component of the measured atom loss (the conversion into shallow dimers and trimers) we have no similar model.
We do note that the total loss rate scaling as $n^{2/3}$ suggests that the Efimov length scale is, at least for this particular combination of observables, much less relevant than the interparticle spacing.
However, the propensity of the system to sweep into shallow molecules appears to depend not only on the sweep rate but on density in a way we don't fully understand yet (see inset of Fig. 2).
We believe the dependence of molecule formation on sweep rate offers information about the length scale of correlations that form in the strongly interacting gas \cite{altman2005dynamic}.
However, preliminary data exploring the multi-dimensional parameter space of $\tau_{\rm{evolve}}$, sweep rates, and even the time at weak interactions have displayed nonseparable dependencies beyond a non-trivial $n$ dependence.

In conclusion, we have created a $^{85}$Rb, $^{85}$Rb$^{*}_2$ and $^{85}$Rb$^{*}_3$ mixture by sweeping a resonantly interacting BEC onto weak interactions.
We believe a better theoretical understanding of how a many-body wave function evolves into the molecular states as interaction strength is lowered may suggest experiments to directly probe the few- and many-body interactions in the resonantly interacting degenerate Bose gas.

The authors thank John Bohn and Francesca Ferlaino for discussions.
Z.H. acknowledges support from EPSRC (Grant No. EP/N011759/1).
This work is supported by the NSF (Grant No. 1125844 and 1607294), NASA, and NIST.

\bibliographystyle{unsrt}
\bibliography{bibv3}

\begin{thebibliography}{10}
\bibitem{comment2}
Tetramer states associated to the first excited Efimov state become unbound for
  values of $a$ larger than our final value and the only existing tetramer
  states are those associated with the Efimov ground state. Such states,
  however, have energies too large and sizes too small compared to $E_n$ and
  $n^{-1/3}$ to be measureabley populated.
\end{thebibliography}

\end{document}